\documentstyle[11pt]{article}
\def\lsim{\mathrel{\rlap{\lower 4pt \hbox{\hskip 1pt $\sim$}}\raise
1pt\hbox{$<$}}}
\def\gsim{\mathrel{\rlap{\lower 4pt \hbox{\hskip 1pt $\sim$}}\raise
1pt\hbox{$>$}}}

\def\cdm{{{\scriptscriptstyle {\rm CDM}}}}

\begin{document}
\begin{titlepage}
\begin{flushright}
SUSSEX-AST 93/4-2 \\
LANCS-TH 93/1\\
(April 1993)\\
\end{flushright}
\begin{center}
\Large
{\bf Inflation and Mixed Dark Matter Models\\}
\vspace{.3in}
\normalsize
\large{Andrew R. Liddle$^*$ and David H. Lyth$^{\dagger}$} \\
\normalsize
\vspace{.6 cm}
{\em  $^*$Astronomy Centre, \\ Division of Physics and Astronomy, \\
University of Sussex, \\ Brighton BN1 9QH.~~~U.~K.}\\
\vspace{.4cm}
{\em $^{\dagger}$School of Physics and Materials, \\ University of
Lancaster, \\ Lancaster LA1 4YB.~~~U.~K.}\\
\end{center}
\vspace{.6 cm}
\baselineskip=24pt
\begin{abstract}
\noindent
Recent large scale structure observations, including COBE, have prompted many
authors to discuss modifications of the standard Cold Dark Matter model.
Two of these, a tilted spectrum and a gravitational wave contribution to COBE,
are at some level demanded by theory under the usual assumption that inflation
generates the primeval perturbations. The third, whose motivation comes by
contrast from observation, is the introduction of a component of hot dark
matter to give the Mixed Dark Matter model. We discuss the implication of
taking these modifications together. Should Mixed Dark Matter prove necessary,
very strong constraints on inflationary models will ensue.
\end{abstract}

\vspace{1cm}
\begin{center}
E-mail addresses:  arl @ uk.ac.sussex.starlink; lyth @ uk.ac.lancs.phy.v1
\end{center}
\end{titlepage}

\section{Introduction: New Parameters for CDM}

Recent observations of large scale structure in the universe, and particularly
that of the Cosmic Microwave Explorer (COBE) satellite DMR experiment (Smoot
{\em et al} 1992), have been widely interpreted as indicating that the
standard Cold Dark Matter (CDM) model gives a qualitatively correct picture of
structure formation, but requires quantitative modifications. Relying on a
gaussian, Harrison--Zel'dovich initial spectrum, the standard CDM model is
specified by a single parameter, the amplitude of the power spectrum, and the
success of this model on confrontation with observation is truly remarkable
(Efstathiou 1990; Liddle \& Lyth 1993). Nevertheless, the required amplitude
as inferred on small scales by pairwise velocities or cluster abundances
appears to differ by a factor of around two from that required by COBE, and
the pattern of clustering in the galaxy distribution on intermediate scales
appears to indicate that the standard CDM spectrum has an incorrect shape on
these scales.

The post-COBE rush of papers has introduced, among other things, three
prominent new parameters into the CDM model. These are\footnote {We are not
considering here the possibility of a fourth parameter, namely the
introduction of a cosmological constant, since it does not have very strong
motivation from either theory or observation (though see Kofman, Gnedin \&
Bahcall 1992).}
(Liddle, Lyth \& Sutherland 1992; Wright {\em et al} 1992; Krauss \& White
1992; Schaefer \& Shafi 1992; Davis, Summers \& Schlegel 1992; Taylor \&
Rowan-Robinson 1992; Cen {\em et al} 1992; Salopek 1992; Liddle \& Lyth 1992;
Adams {\em et al} 1993)
\begin{itemize}
\item A gravitational wave component contributing a fraction $R$ of the large
angle microwave anisotropies.
\item A `tilt' $n$ of the primeval spectrum away from a flat ($n=1$) spectrum.
\item An admixture of hot dark matter (HDM), contributing a fraction
$\Omega_{\nu}$ to the critical density
\end{itemize}
To our knowledge, all three have not been considered together before this
paper. A hot dark matter component has most often been discussed without
either of the others, though Schaefer and Shafi (1993) have included tilt in
their studies but not gravitational waves. Particularly in $N$-body studies,
tilt is typically discussed on its own, but the present authors have offered a
study including both tilt and gravitational waves (Liddle \& Lyth 1993). One
aim of this paper is to examine the rationale behind these new parameters,
both from a theoretical and observational viewpoint.

The motivation for these new parameters differs. A component of hot dark
matter is clearly an optional extra. In order to have a third component of
nonrelativistic matter (along with the baryonic and cold dark matter
components) with density of the order of the critical density, it appears that
some form of tuning of the parameters is needed. The relative abundances of
particle species is a complicated function of masses and couplings, and can in
principle take on a wide range of values. To have CDM and baryons with similar
densities is already a modest coincidence; to have the HDM and baryon density
similar too exacerbates this. Nevertheless there may well exist particle
physics models which do exhibit these appropriate tunings. The point to make
here is that this extra parameter is at best poorly motivated by presently
understood particle physics; the reason why the so-called mixed dark matter
(MDM) model\footnote{Often denoted C+HDM (or CPHDM), which is more descriptive
but more cumbersome.}, where $\Omega_{\nu} \simeq 25$--$30$\%, is so popular is
because of its strong phenomenology, as we shall discuss.

The other new parameters are not just well founded theoretically --- provided
one takes the step of believing inflation as the cause of the primeval
inhomogeneities they are at some level inevitable. The inflationary prediction
has long been advertised as a Harrison--Zel'dovich spectrum with a
gravitational wave spectrum of negligible amplitude (Kolb \& Turner 1990;
Linde 1990). In the past, when observations were restricted to a limited range
of scales, this was a very reasonable approximation. However, with COBE
probing length scales vastly in excess of those studied previously, it seems
that this approximation is no longer good enough (Davis {\em et al} 1992;
Salopek 1992; Liddle \& Lyth 1992; Adams {\em et al} 1993). The generic
prediction from inflation is a primeval density perturbation spectrum that can
be well approximated by a power-law $P(k) \propto k^n$, but where $n$ is
`tilted' from the flat $n=1$ case. Almost generically, the tilt is to $n<1$,
removing short-scale power from a COBE normalised spectrum\footnote{It is
possible to have $n>1$, but this is not achieved by any of the well-accepted
inflationary models and with present understanding should be regarded as
unnatural.}. Gravitational waves are also generically created, and though
their contribution to COBE is normally less than that of the density
perturbations, it can easily be tens of percent, which can in no way be
regarded as negligible.

A case in point is provided by the simplest model of inflation, chaotic
inflation with a free massive scalar field (Linde 1990). This model produces
the {\em smallest} distortive effects amongst the more popular inflationary
models. The deviation from the flat spectrum is technically logarithmic, but
in practice excellently described by a power-law over scales of interest. The
combined effects of tilt and gravitational waves in this model reduce
$\sigma_8$, the dispersion of the density field at $8h^{-1}$ Mpc, by 13\%
(when normalised to the COBE $10^0$ result). This is not startling, but it
is the minimum deviation expected from inflation and certainly large enough to
convert an unlikely looking $2$-sigma result into a comfortable $1$-sigma
result at present observational accuracy. In a self-coupled chaotic inflation
model, this leaps to 20\%, and in others yet more.

While the complete details of predicting the perturbation spectra from
inflation are quite involved, the outcome is very simple. In present versions,
an inflationary model consists of no more than a scalar field $\phi$ evolving
in a potential $V(\phi)$ and a mechanism to end inflation. The form of the
potential is largely up for grabs, but barring pathologies the following
parameters, the slow-roll parameters,  must be small compared with unity,
\begin{eqnarray}
\epsilon & = & \frac{m_{Pl}^2}{16\pi} \left( \frac{V'}{V} \right)^2\\
\eta & = & \frac{m_{Pl}^2}{8\pi} \frac{V''}{V}
\end{eqnarray}
where primes are derivatives with respect to $\phi$. In general these depend
on the scalar field value $\phi$, but usually the scales of interest for large
scale structure leave the horizon over a short interval and they can be
treated as constant. The exception is `designer' models of inflation, where
one contrives dramatic features in the potential just at the points
appropriate for large scale structure. These simple parameters are vitally
important, because {\em they alone determine to high accuracy\footnote{It
has recently been verified by explicit calculation that the corrections to the
following formulae are indeed small (Stewart \& Lyth 1993).} the degrees of
tilt of both the density perturbation and the gravitational waves, as well as
their relative normalisations at the COBE scale} (Davis {\em et al} 1992;
Liddle \& Lyth 1992).

Scales of cosmological interest leave the horizon about 60 $e$-foldings from
the end of inflation (that is, when the scale factor was smaller by a factor
$e^{60}$ than at the end), and it is the value of the slow-roll parameters
then that is required. Usually, inflation ends when the field approaches a
minimum, and $\epsilon$ exceeds unity\footnote{There are exceptions, where
more involved means of ending inflation are introduced. The key example is
power-law inflation, which requires an exponential potential. Conveniently,
the slow-roll parameters are exactly constant in this case, so the prediction
for the spectra is independent of the precise means of ending inflation. Other
exceptions are discussed by Liddle \& Lyth (1993).}. The scalar field
value $N$ $e$-foldings from the end of inflation is easily obtained via
\begin{equation}
N = - \frac{8\pi}{m_{Pl}^2} \int_{\phi}^{\phi_e} \frac{V}{V'} \; {\rm d}\phi
\end{equation}
where $\phi_e$ is the value at the end of inflation. Given a potential, it is
thus easy to calculate the appropriate scalar field value, and hence the
parameters $\epsilon$ and $\eta$.

The degree of tilt of the density perturbation, defined as the departure of
its spectral index from the scale invariant value $n=1$, is
\begin{equation}
1 - n = 6\epsilon - 2 \eta
\end{equation}
The degree of tilt of the gravitational wave amplitude, defined as the
departure of its spectral index from the scale invariant value $n_g=0$, is
\begin{equation}
-n_g=2\epsilon
\end{equation}
Finally, the ratio $R$ of the gravitational wave and density perturbation
contributions to the expected mean square microwave background anisotropy
measured by COBE is
\begin{equation}
R \simeq  12\epsilon
\end{equation}
The tilt of the gravitational wave amplitude will be hard to measure because
it can be probed only through the microwave background, but the actual
magnitude
could be crucial. In plain language, if gravitational waves are significant
the {\em rms} density perturbation $\sigma$ when normalised to COBE is only a
fraction $F = 1/\sqrt{1+R}$ of what you thought it would be.

For example, consider the potential $V(\phi) \propto \phi^{\alpha}$, proposed
in the context of chaotic inflation (Linde 1990). From above, $\phi_{60}/m_{Pl}
\simeq \sqrt{60 \alpha/4\pi}$, so
\begin{equation}
\epsilon = \frac{\alpha}{240} \quad ; \quad \eta= \frac{\alpha - 1}{120}
\end{equation}
So we immediately know that $n = 1 - (2+\alpha)/120$, and that the
gravitational waves will reduce the dispersion $\sigma_8$ by a factor
$1/\sqrt{1+\alpha/20}$. For the smallest conceivable power $\alpha=2$
(corresponding to a free field), the tilt can be shown to normalise $\sigma_8$
down by 8\% relative to CDM, and the gravitational waves by a further 5\%, as
advertised above.

In this polynomial model the degree of tilt $1-n$ and the relative
contribution of the gravitational waves $R$ are related by
$6(1-n)=0.1 +R$. A similar relation, $6(1-n)=R$, holds
in the power-law inflation models mentioned
earlier, but such a relation is not generic and in particular one can have
significant tilt without significant gravitational waves (Liddle \& Lyth 1992;
Adams {\em et al} 1993). The essential point, though, is that barring
fine-tuned `designer' models the whole gamut of possible inflationary models
introduces only two additional parameters into the standard CDM model.

\section{Observations}

The two parameters associated with generic inflation models, plus the third
one invoked by the MDM model, have different effects. Let us assume for the
time being that the theory is normalised to the COBE observations. The
gravitational waves simply normalise down the amplitude of the whole spectrum.
Tilt removes short-scale power from the spectrum, progressively across the
whole range of scales. The hot dark matter component, on the other hand,
removes power from the spectrum only up to the scale on which free-streaming
of the HDM can occur, typically tens of megaparsecs, while leaving the large
scales the same as in CDM. The effect of the HDM is somewhat subtle, in that
usually only one free parameter, $\Omega_{\nu}$, is allowed. It is then assumed
that the HDM has standard properties, effectively those of a neutrino, which
relate its abundance to its mass\footnote{Should things go badly for MDM,
there is a rather unpalatable opportunity to add an extra parameter here,
measuring in some way the `nonstandardness' of the connection between the
relic abundance and mass of the HDM particles.}. The mass then finally
provides the free-streaming length. The one parameter thus determines both the
extent to which free-streaming removes short-scale power, and also the scale
up to which the free-streaming is effective. The success of the MDM model is
that with a choice of the parameter as $25$--$30$\%, these two features are
respectively of the size, and at the scale, at which one would wish them,
allowing one free parameter to simultaneously fit several pieces of data.

One should note that, barring unusual inflation, all the new parameters serve
to progressively subtract power relative to the CDM spectrum as one progresses
to shorter scales. Observations on a given scale are commonly interpreted by
giving the amplitude a CDM spectrum would require to explain them; this is
normally specified by the $\sigma_8$ this amplitude would give, regardless of
the scale on which the observations apply. As long as observations towards
progressively smaller scales lead to progressively smaller predictions for the
CDM amplitude, then one can reasonably expect these new parameters to be
useful. It so happens that the available observational data is precisely of
this sort.

There are a variety of observations giving the amplitude of the mass
fluctuations across a range of scales. However, it appears possible to take a
very crude view and conclude that there are effectively only about four
different measurements which one must satisfy, as different measurements on
the same scales appear pretty much in agreement. We list such a set below,
noting in each case the value of $\sigma_8$ which would be required in the
standard CDM model, denoted by $\sigma_8^\cdm$.
\begin{itemize}
\item Scales $10^3 h^{-1}$ to $10^4h^{-1}$ Mpc\footnote{As usual $h$ is the
Hubble parameter in units of $100\mbox{\,km\,s}^{-1}\mbox{\,Mpc}^{-1}$, and
in making predictions is taken to be equal to $0.5$.}: COBE provides a
measurement of the spectrum here, allowing a one-sigma range from about
$\sigma_8^\cdm = 1$ to $1.3$. The top of this range is however disfavoured by
subsequent analysis and other experiments, and the most likely true value in
the light of these is perhaps the COBE $1$-sigma lower limit.
\item Scales $20h^{-1}$ to $40h^{-1}$ Mpc: Velocity flows appear the most
unambiguous measure here. For example, QDOT (Kaiser {\it et al} 1991) provides
$\sigma_8^\cdm$ ranging from between about $1$ and $0.7$. A comparison of
POTENT with the 1.2 Jansky survey (Dekel {\it et al} 1992) yields similar
results, as does a direct comparison with POTENT bulk flows. Following
Efstathiou, Bond and White (1992), we utilise the QDOT results on the IRAS
bias and number count variance in $30h^{-1}$ Mpc cubes, converted to spheres
of equal volume, {\it ie} radius $19 h^{-1}$ Mpc. Dealing directly with the
variance on this scale, this gives a $1$-sigma range $\sigma_{19} = 0.37 \pm
0.07$. [The prediction of standard CDM at COBE normalisation is $\sigma_{19} =
0.45$, which can be translated into the $\sigma_8^\cdm$ limits above.]
Primarily we are interested in the lower limit, and it is worth noting that
POTENT/IRAS gives a much stronger version, as they obtain a 95\% confidence
upper limit on the IRAS bias which is less than the QDOT $1$-sigma upper
limit.
\item The scale of order $10h^{-1}$ Mpc: A direct measure of the amplitude at
$8 h^{-1}$ Mpc is provided by the abundance of galaxy clusters.  A recent
analysis by White, Efstathiou \& Frenk (1993) gives $\sigma_8^\cdm$ in the
range $0.30$--$0.63$ (the lower limit takes into account the possibility that
current cluster mass estimates could be a factor of 3 or so too big).
Recently, it has been claimed that through non-linear effects the pairwise
galaxy velocity dispersion for galaxy separations of of order $1$ Mpc also
measures the primeval amplitude on the scale of order $10h^{-1}$ Mpc (Gelb,
Gradwohl \& Frieman 1993), requiring roughly $.3\lsim \sigma_8^\cdm\lsim .5$.
We shall use the galaxy cluster range in what follows.
\item The scale of order $1h^{-1}$  Mpc: This comoving scale encloses (before
gravitational collapse) a mass comparable to that of a large galaxy or quasar.
At high redshift such objects are rare because only a small fraction of the
matter has had time to collapse, but lower limits on their abundances can be
estimated which translate into lower limits on $\sigma$ on this scale
(Efstathiou \& Rees 1988; Adams {\em et al} 1993; Cen {\em et al} 1993;
Haehnelt 1993). Haehnelt uses the observed quasar luminosity function (Irwin,
McMahon \& Hazard 1991; Boyle {\em et al} 1991), together with reasonable
assumptions about quasar astrophysics, to deduce that at redshift 4 the
fraction $f(>M)$ of mass bound into objects with $M>10^{13}M_\odot$ is at
least $1\times 10^{-7}$. He compares this result with the Press--Schechter
formula
\begin{equation}
f(>M)=1-\mbox{erfc}\left(\frac{\delta_c}{\sqrt2 \sigma(M,z)}\right)
\end{equation}
where $\delta_c$ is the value {\em in linear theory} of the density contrast
at which gravitational collapse is assumed to take place. Comparisons with
numerical simulations have suggested values of $\delta_c$ in the range 1.3 to
1.7, and taking the lower value one finds a bound on
$\sigma(10^{13}M_{\odot},4) $ which is equivalent to $\sigma_8^\cdm>.40$.
(Haehnelt obtains a somewhat stronger constraint by taking the higher value.)
Note that when comparing this $z=4$ result with the MDM model one has to allow
for the $z$-dependence of the MDM transfer function, reflecting the slower
growth of the perturbation relative to the CDM model.
\end{itemize}

In our view, a model which manages to satisfy all of these limits has a good
chance of agreeing with all other available observations. In particular,
though we have not mentioned it explicitly, it should satisfy all the
clustering data such as the APM survey (Maddox {\it et al} 1990)
and a host of later surveys. When first produced, these were seen, rightly, as
a major problem for CDM, indicating that the clustering strength falls off
less rapidly than expected with increasing scale. However, if one fits the
data above there is clearly going to be an excess in amplitude as one goes
from $8h^{-1}$ to the scales around $20h^{-1}$ Mpc on which the velocity flow
data operates. Conveniently, the excess clustering data has already been
reinterpreted by Wright {\it et al} (1992), using a quantity they call the
`excess power' $E$, defined simply as
\begin{equation}
E = 3.4 \frac{\sigma(25h^{-1} {\rm Mpc})}{\sigma(8h^{-1} {\rm Mpc}}
\end{equation}
The prefactor is chosen to make the CDM value unity, and they suggest that
values of $E$ in the range $1.15$ to $1.45$ will fit the clustering data. We
see from the figures above that if we fit the cluster abundance and velocity
flow data, we can hardly fail to satisfy this excess power
criterion\footnote{We have specifically mentioned APM. However, other surveys
indicating excess power appear consistent with APM --- this is normally
indicated by authors quoting viable power spectra in terms of the $\Gamma =
\Omega h$ parametrisation of Efstathiou, Bond \& White (1992). Usually $\Gamma
= 0.2$--$0.3$ is required (Kofman {\it et al} 1992), which can be translated
into the same excess power criterion.}.

\section{Confrontation with observations}

The figures illustrate the parameter space regions which satisfy these data
points. This is intended only to be illustrative of trends, for two reasons.
Firstly, the observational data are not particularly strict, and most people
would accept a reasonable amount of freedom to manipulate the figures above.
Secondly, the theoretical calculations are not as accurate as one would like.
We have utilised transfer functions from van Dalen and Schaefer (1992), who
supply parametrised forms for a set of $\Omega_{\nu}$, to make our
calculations. However, these are only accurate to perhaps ten percent or worse
across the full range of scales, which in many places is comparable to the
observational uncertainties. Klypin {\it et al} (1992) have provided what
appears a more accurate transfer function for $\Omega_{\nu} = 0.30$; when
normalised to COBE it gives values for the dispersion of between 10\% and 15\%
lower across the scales where we compare with data.

Equally, one should not be expecting any dramatic conclusions. After all, we
are allowing four free parameters (amplitude, tilt, gravitational waves, HDM
fraction) and have only four data points to fit. So what are we looking for?
The important points appear to be the following
\begin{enumerate}
\item The three parameters naturally motivated by inflation are amplitude,
tilt and gravitational waves. Given such freedom, it is perhaps surprising
that they appear insufficient to allow one to fit the data (see Liddle \&
Lyth (1993) for a more extensive discussion). Because of its progression
across the entire range of scales, tilt seems incapable of providing the sharp
drop in power between the bulk flow and cluster abundance scales. As a further
symptom of the same shortcoming, it fails to fit the APM data if it fits
QDOT. A modification, such as MDM, seems essential.
\item As advertised, satisfying the cluster abundance and bulk flow data
effectively guarantees a fit to clustering data such as APM.
\item MDM without tilt or gravitational waves does rather well. Unfortunately,
one cannot motivate the complete absence of tilt and gravitational waves by
appealing to inflation. So it appears that one really ought to allow all three
new parameters.
\item MDM only works well provided that the tilt and gravitational waves are
very small. Thus, if one believes the MDM model one has to accept very strong
constraints on models of inflation.
\end{enumerate}
The last point is worthy of additional comment. Our present understanding of
the fundamental interactions, which is embodied in the Standard Model, does
not lead to inflation. However, the Standard Model has been tested only on
energy scales $\lsim100$ GeV, and for reasons that have nothing to do with
cosmology one anticipates an extension of the Standard Model at higher energy
scales. The subject of particle cosmology came into being when it was realised
that the early universe constitutes a `cosmic accelerator', allowing one to
probe energy scales far beyond the reach of laboratory physics, and the last
point is a specific example of this remarkable fact. In another publication
(Liddle \& Lyth 1992) we have already ruled out a class of otherwise
attractive inflationary models (if they are to generate the density
perturbation required to form structure), and we are here pointing to the
possibility of drawing stronger conclusions from better data.

As an illustration of the sort of thing that might become possible, let us
suppose that the observational bounds all turn out to be correct, and that MDM
turns out to be necessary. Then with COBE normalisation, the degree of tilt
$1-n$ is constrained to the range 0.07 to 0.10 in models with negligible
gravitational waves, and to the range 0.03 to 0.05 for chaotic and power-law
models. For the former case, the exponent $\alpha$ in the potential is
constrained to the range 1.6 to 4.0, which would for example rule out the
otherwise attractive inflationary model recently proposed by Lazarides
and Shafi (1993)
within the context of superstrings. While it should not be taken seriously at
the present time, this example serves as a reminder that every improvement in
observational cosmology has potential implications regarding the search for a
viable model of the fundamental interactions.

\section*{Acknowledgements}
ARL is supported by the SERC, and acknowledges the use of the Starlink
computer system at Sussex. We thank Martin Haehnelt and Qasir Shafi sending us
the preprints mentioned in the text, and Jon Holtzman and Joel Primack for
correspondence about the MDM transfer function.

\section*{References}
\frenchspacing
\begin{description}
\item Adams, F. C., Bond, J. R., Freese, K., Frieman, J. A., Olinto, A.
	V., 1993, Phys. Rev. D47, 426
\item Boyle, B. J., Jones, L. R., Shanks, T., Marano, B., Zitelli, V.,
	Zamorani, G., 1991, in Crampton, D., ed., ASP Conference
	Series No. 21, {\sl The Space Distribution of Quasars},
	page 91 (Astronomy Society of the Pacific, San Francisco)
\item Cen, R., Gnedin, N. Y., Kofman, L. A., Ostriker, J. P., 1992,
	ApJ, 399, L11
\item Davis, M., Summers, F. J., Schlegel, D., 1992, Nat, 359, 393
\item Davis, R. L., Hodges, H. M., Smoot, G. F., Steinhardt, P. J.,
	Turner, M. S., 1992, Phys. Rev. Lett., 69, 1856
 \item Dekel, A., Bertschinger, E., Yahil, A., Strauss, M. A., Davis, M.,
	Huchra, J. P., 1992, ``IRAS Galaxies verses POTENT mass: Density
	Fields, Biasing and $\Omega$'', Princeton preprint IASSNS-AST 92/55
\item Efstathiou, G. 1990, in ``The Physics of the Early Universe'', eds
	Heavens, A., Peacock, J., Davies, A., SUSSP publications
\item Efstathiou, G., Rees, M. J., 1988, MNRAS, 230, 5p
\item Efstathiou, G., Bond, J. R., White, S. D. M., 1992, MNRAS, 258, 1p
\item Gelb, J. M., Gradwohl, B.-A., Frieman, J. A., 1993, ApJ, 403, L5
\item Haehnelt, M. G., 1993, ``High redshift constraints on alternative spectra
	for primeval density perturbations'', Cambridge Institute of Astronomy
	preprint
\item Holtzman, J., Primack, J., 1993, ApJ, 405, 428
\item Krauss, L. M., White, M., 1992, Phys. Rev. Lett., 69, 869
\item Irwin, M., McMahon, R. G., Hazard, C., 1991, in Crampton, D., ed.,
	ASP Conference Series No. 21, {\sl The Space Distribution of Quasars},
	page 91 (Astronomy Society of the Pacific, San Francisco)
\item Kaiser, N., Efstathiou, G., Ellis, R., Frenk, C., Lawrence, A.,
	Rowan-Robinson, M., Saunders, W., 1991, MNRAS, 252, 1.
\item Kofman, L., Gnedin, N., Bahcall, N. A., 1992, ``Cosmological
	Constant, COBE CMB Anisotropy and Large Scale Clustering'', CITA
	preprint 92/94
\item Klypin, A., Holtzman, J., Primack, J. R., Reg\"{o}s, E., 1992,
	``Structure Formation with Cold + Hot Dark Matter'', Santa Cruz
	preprint SCIPP 92/52
\item Kolb, E. W., Turner, M. S., 1990, \sl The Early Universe
	\rm (Addison-Wesley)
\item Lazarides, G. and Shafi, Q.
``A Predictive Inflationary Scenario without the
	Gauge Singlet'', Bartol preprint
\item Liddle, A. R., Lyth, D. H., 1992, Phys. Lett. B291, 391
\item Liddle, A. R., Lyth, D. H., 1993, Phys. Rep., to appear
\item Liddle, A. R., Lyth, D. H., Sutherland, W., 1992, Phys. Lett. B279,
	244
\item Linde, A. D., 1990, \sl Particle Physics and Cosmology \rm
	(Gordon and Breach)
\item Maddox, S. J., Efstathiou, G., Sutherland, W. J., Loveday, J., 1990,
	MNRAS, 242, 43p
\item Salopek, D. S., 1992, Phys. Rev. Lett., 69, 3602
\item Schaefer, R. K., Shafi, Q., 1992, Nat, 359, 199
\item Schaefer, R. K., Shafi, Q., 1993, Phys. Rev., D47, 1333
\item Smoot, G. F. {\it et al}, 1992, ApJ, 396, L1
\item Stewart, E. D., Lyth, D. H., 1993, Phys. Lett., B302, 171
\item Taylor, A. N., Rowan-Robinson, M., 1992, Nat, 359, 396
\item van Dalen, T., Schaefer, R. K. 1992, ApJ, 398, 33
\item White, S. D. M., Efstathiou, G., Frenk, C. S., 1993, ``The
	Amplitude of Mass Fluctuations in the Universe'', Durham preprint
\item Wright, E. L. {\it et al}, 1992, ApJ, 396, L13

\end{description}
\nonfrenchspacing

\vspace{24pt}
\section*{Figure Captions}

\vspace{0.5cm}

{\em Figure 1}\\
The constraints in the $n$--$\Omega_{\nu}$ parameter space, for (a),
inflationary models with no gravitational waves, COBE normalised; (b),
chaotic (or power-law) inflationary models, incorporating gravitational waves,
COBE normalised; and (c), as (b), but normalised to the COBE $1$-sigma lower
limit. The lines shown are solid, quasar abundance; dashed, cluster abundance;
dot-dashed, bulk flows from QDOT; dotted, clustering data from APM. The shaded
region indicates the region satisfying all data (and can be used in each case
to see which side of the line is the allowed side). We strongly urge the reader
to treat the details with skepticism, following the caveats in the text, but
to pay attention to the trends and possibilities. Finally, the notches on the
top axis of (b) and (c) indicate the location of chaotic inflation models with
exponent $\alpha$=2, 4, 6 and 8.
\end{document}